\documentclass[runningheads]{llncs}
\usepackage{graphicx}

\usepackage[T2A]{fontenc}

\begin{document}

\title{Detecting Automatically Managed Accounts in Online Social Networks: Graph Embedding Approach}
\author{Ilia Karpov \orcidID{0000-0002-8106-9426} \and
Ekaterina Glazkova\orcidID{0000-0001-5435-4064}}

\authorrunning{Ilia Karpov \and Ekaterina Glazkova}
\titlerunning{Detecting Automatically Managed Accounts in OSN}
\institute{National Research University Higher School of Economics, Moscow, Russian Federation
\email{\{karpovilia, catherine.glazkova\}@gmail.com}}

\maketitle

\begin{abstract}
The widespread of Online Social Networks and the opportunity to commercialize popular accounts have attracted a large number of automated programs, known as artificial accounts. This paper\footnote{Project repository available at \url{http://github.com/karpovilia/botdetection}} focuses on the classification of human and fake accounts on the social network, by employing several graph neural networks, to efficiently encode attributes and network graph features of the account. Our work uses both network structure and attributes to distinguish human and artificial accounts and compares attributed and traditional graph embeddings. Separating complex, human-like artificial accounts into a standalone task demonstrates significant limitations of profile-based algorithms for bot detection and shows efficiency of network structure based methods for detecting sophisticated bot accounts. Experiments show that our approach can achieve competitive performance compared with existing state-of-the-art bot detection systems with only network-driven features.

\end{abstract}

\section{Introduction}
The need to quickly search for artificially created accounts is in many practical research tasks. Artificial accounts distort the popularity of groups \cite{candidatepopularity}, spread fake news \cite{Ratkiewicz}, \cite{shao2017spread}, and are used for fraud \cite{Apte2019} activities. In this work we define an artificial account as an account created and used to generate profit for the owner by violating the rules of a social network. Thus, intuitively, accounts can be divided into two types:

\begin{itemize}
    \item technical accounts (or bots) created to collect data, increase the number of group members, the popularity of posts etc. Most of these accounts are created and controlled by software. As a rule, these are poorly filled and relatively recently created profiles, which allows to identify them with a sufficiently high accuracy using several profile features.
	\item sophisticated accounts, created and operated semiautomatically by a human, are used primarily for information propagation, marketing and fraud activity. This category also includes hacked and used for fraud accounts.
\end{itemize}

Most of the existing methods show excellent results in identifying accounts of the first type. However, they do not consider accounts of the second type due to the high complexity of dataset acquisition. At the same time, accounts of the second type are visually and in terms of profile fullness indistinguishable from legitimate user accounts, which makes the task of their classification much more challenging.

The problem of finding artificial accounts of the second type can be reduced to the task of classifying network nodes. Still, its solution requires much more contextual information and learning capacity of the models used than in case of general node classification problem. This makes it appropriate to study the applicability of existing models for describing network nodes and their classification to solve this problem.

This paper has two main contributions as follows:

\begin{itemize}
    \item First, we introduce the two bot detection dataset, consisting of nodes of three types: real user accounts, technical accounts and manual accounts. Unlike many existing arrays, it contains not only information about the network profile, but also information about the user's friends and groups. Since the array contains data of real people, immediately after collecting the data, all identifiers of users, groups, places of residence were hashed. An important advantage of the proposed array is that it does not contain precalculated walk models, but the initial hashed data themselves for training various models. Thus, it can be used in the future to train and compare new models.
    \item Second, we investigate two models for constructing network embeddings to solve the problem of classifying network nodes into artificial and natural. As far as we know, we are the first who use a class of the network node attributes simultaneously with the network node embedding to solve this problem.
\end{itemize}

The use of random walk algorithms optimized for large networks is also of interest. This comparison is especially important given that the problem of identifying real life bots is solved for a graph of
\( \sim10^9 \) nodes, which imposes significant restrictions on the computational cost of the algorithms used.

The rest of the work is organized as follows. In Section 2 we describe existing approaches to bot detection problem and recent advances in network embedding generation. Section 3 is devoted to the proposed model. Section 4 presents our experiments. At the Section 5 we summarize experiment results and discuss future work.

\section{Related work}
This section describes the chosen vector representations with attributes approach, the formal criteria to separate simple bots from advanced ones.

\subsection{Labeled datasets}

Due to the lack of an established definition of the "bot" term, as well as the illegality of the very usage of bots, the formation of a training dataset causes problems. Based on the definition of the above and previously performed works \cite{morstatter}, \cite{ispras}, the main methods are (1) manual bot labeling - this approach is very resource-intensive and does not work well for sophisticated bots due to the low annotators agreement coefficient, (2) monitoring of suspended users lists - the main disadvantage of this approach are that (a) it is necessary to have time to proactively collect the entire user's network profile before blocking, (b) not all users blocked by a social network due to rules violation are bots (3) Creating accounts that attract bots other than real people, this approach looks very original, but most likely attracts bots using certain communication strategies, which does not fully give an idea of the diversity.

In this work, we use the second approach mainly to form a set of technical bots along with one more previously undescribed strategy. To construct a set of sophisticated accounts, we got access to \(\sim 700\) accounts sold by 12 different sellers on special account exchanges. The purchase of accounts is prohibited by the social network rules, all scenarios for using purchased accounts are negative and stay within the proposed definition, which makes monitoring of account exchanges a fairly effective strategy for generating the dataset. At the same time, advanced accounts, as a rule, have a higher price, which allows constructing a cost scale that correlates with the difficulty of identifying such an account. With it, the part of suspended users from the first set can be attributed to sophisticated accounts using the k-nearest neighbors’ method.

\subsection{Existing approaches}
Researchers designed many bot detection approaches in order to prevent the spread of social media bots. Those approaches can be classified into 3 main categories: graph-based, feature-based and anomaly-based. 

The anomaly-based approaches assume that a user with odd behaviour is most likely to be a malicious user. For example \cite{anomaly1} took into assumption that human behaviors exhibit strong heterogeneity and a bot’s behavior is less complex than a human’s behavior which leads to a smaller entropy of bot actions. They analyse time-interval entropy value and clustered the accounts on Sina Weibo into humans, bots and Cyborgs. Another work \cite{anomaly2} created empty accounts and managed to construct a bot dataset with the users that interacted with these empty accounts.

There are several feature-based methods where authors identified a set of behaviour features which describes user actions, interactions, timestamps and may include text counting without deeply analyzing textual content. Then they train several supervised machine learning classifiers like SVM-NN in \cite{sup_svm_nn} or Random Forest Classifier in \cite{sup_rfc}. The sequence of user's actions can also be used as input to CNN-LSTM algorithm \cite{lstm}

Besides behaviour features researchers use content-based features such as tweets. For example in \cite{sentiment} authors use sentiment analysis and build Contrast Pattern-Based classifier. Besides tweets user's nicknames  also can be analysed. Thus, \cite{nicks} classified user names (strings) into random and non-random and it can be used as additional feature for bot detection classifiers.

Graph-based bot detection methods rely on analyzing the structure of social network. The majority of them make the assumptions that real users form one strongly connected community and refuse to interact with artificial users \cite{SybilGuard},\cite{SybilLimit}, \cite{SybilInfer}. These assumptions mean that there is a sparse cut between the legitimate and bot parts. \cite{ispras} describe an approach to bot detection using a stacking ensemble with classifier trained on top of a combination of node embeddings from friendship graph, models trained on user's text and bag of subscribers. They use graph embedding obtained from LINE while in our work we compare various different graph embeddings.

\subsection{Graph embeddings}
Given a weighted graph \textit{G = (V, E, A)}, where \textit{V} is the set of nodes ,\textit{E} - is the edge set and \textit{A} is the adjacency matrix of a graph \textit{G} such as \(A_{ij} = weight(v_i,v_j)\) if \((v_i, v_j ) \in E \), otherwise \(A_{ij} = 0\), the goal is to learn a function \textit{\(V \rightarrow R^d\)} that that maps each vertex to a d-dimensional \textit{ \((d << |V|)\)} vector that captures its network connectivity structure. Resulting embedding can be used as input vector for the classification problem.

After the successful application of skip-gram \cite{mikolov} model to the problems of modeling text sequences, this method began to be actively used to model the graph nodes based on linear sequences obtained as a result of the graph random walks. The main advantage of using neural network methods in graph problems is that they do not need the whole training set during the training phase. They optimize by reading the data by batches step by step and correcting continuously which theoretically leads to training on extremely large datasets. In practice most of the frameworks require precomputed transition probabilities, that should be stored in memory.

Attri2Vec \cite{attri2vec} generates random walks to capture structural context. After random walks are generated, attri2vec uses the representation \(f(x_i)\) of node \(v_i\) with attribute \(x_i\) in the new attribute subspace to predict its context nodes collected from random walk sequences. In this way, network structure is seamlessly encoded into the new attribute subspace by allowing nodes sharing similar neighbors to be located closely to each other.

Node2Vec \cite{node2vec} is the classical second-order random walk based on the transition parameters \(p\) and \(q\), where \(p\) controls the probability to return back to the starting node of the walk and \(q\) - to walk away from the starting node of the walk. Given \( p = 1 \)  and \(q = 1 \)  the algorithm models unparameterized random walk.

\section{Proposed Approach}
VKontakte Social Network provides a significant amount of information about the user that can be converted into features. The features can be divided into five groups:
\begin{itemize}
    \item Descriptive profile characteristics. These features include gender, age, marital status, date of birth, number of friends, groups, subscribers, country and city of residence, school, university, user work, site, number of posts on the user's wall, number of photos and albums, audio and video records, privacy settings, confirmation by phone number, date of account creation; geotags of photos, links to profiles in other networks;
    \item Text features received from user's messages, comments and statuses. They can include text embeddings, key topics of messages and comments, variety of vocabulary, user language;
    \item Image features, images on the page and in the user's albums;
    \item Graph features - user group membership, user friends structure;
    \item Time features such as user last seen time, publication time, activity at neighbor pages such as likes, comments at friends and subscribed groups;
\end{itemize}

In this work, we do not use text, image and time features due to the desire to focus on graph features. We still download text and profile information for further research and dataset enrichment.

\subsection{Dataset generation}

In our work we consider two sources of artificial accounts information: blocked by the VKontakte social network administration ("banned") and placed on a special exchange for the purpose of selling ("sold"). All accounts have one of the following statuses: active, deleted and banned. We consider only accounts with status banned in "banned" list, which excludes users who have independently deleted their accounts from the network from being included in the set ones. This approach has a certain error ratio, since, as noted above, not only accounts participating in increasing the rating of groups and disseminating information are subject to blocking, but also real users blocked by the administration, for example, due to incorrect behavior towards other participants or publications of prohibited materials. We suggest focusing on "sold" accounts to train more precise models.

For obtaining "banned" accounts we used the following procedure:
\begin{enumerate}
    \item Users profiles with friends and texts were collected during February 2020. Only profiles, that are accessible without the social network registration were collected.
    \item In June 2020 profiles from step 1 were checked with friends and friends of friends. Two lists of users with statuses "active" and "deactivated - banned" are collected.
    \item Graph based on found users, their connections and their friends and friends of friends was created. 
    \item The largest connected component of the obtained graph is considered.
\end{enumerate}

When constructing the final graph, we examined 716 sophisticated and 360 technical accounts found from account shops and 2515 blocked users before they were banned. Blocked users were classified as"technical" and "sophisticated" based on their profile fulfillment and friends count. We obtained mutual friends for all nodes (bots and their friends) for the largest connected component, that gave us 322,917 nodes totally. The resulting connected graph consists of 270 technical accounts and 159 sophisticated accounts and 44,667 users with all information about profile, groups and friends; This node framing strategy distorts periphery nodes centrality measures, since their edges with nodes outside the frame that are not taken into account. It can cause an errors in the results, since it is bot egocentric, but a full-rated modeling of the second-order friends graph is complicated by the fact that it contains about 12 million nodes, which does not allow performing random walks in an acceptable time.

Sophisticated and technical accounts were represented in the largest connected component, which made it possible to train one graph embedding model to classify both types of accounts.

For "banned" users, an additional assessment of the profile filling in was carried out based on the profile filling in by such features as the number of friends, subscribers, groups, photo, account verification.

\subsection{Selected Features}
To train the basic model, we used the following account attributes: age, {account verification by phone, nickname , website, Facebook profile, Instagram profile, Twitter profile, photo} flags, number of subscriptions to public pages, number of videos, number of audio, number of days since the last login to the network, filled status, number of friends, city, gender.

In the case when individual fields were not filled in or were not available for collection, we substituted the median value. In the future, it seems appropriate to fill in these fields as the median value for k nearest neighbors of the user or to apply more advanced approaches to filling described in the work \cite{ferligoj}.

To train graph embedding, the user's friend graph was used. Features of encoder walking and training will be described in the section below.

\subsection{Graph embeddings}
We considered two types of embeddings: Node2Vec \cite{node2vec} and Attri2Vec \cite{attri2vec}. To obtain Node2Vec embeddings for our graph we used the SNAP framework \cite{snap}. For Attri2Vec embeddings we used Attri2Vec authors implementation \footnote{{\url{https://github.com/daokunzhang/attri2vec}}}. More implementation details are presented in section Experiments.

\subsection{Classification}
The resulting graph embedding features were combined with the user feature vector and used as input parameters of the classification algorithm. The following classification algorithms were considered:

\begin{enumerate}
    \item Support Vector Classifier (SVC)
    \item Random Forest (RF)
    \item Logistic Regression (LogReg)
\end{enumerate}

Implementation of algorithms from scikit-learn \cite{scikit} was used.

\section{Experiment}

\subsection{Classification on account features}

Logistic regression classification algorithm was trained based on the accounts features. Features described in Section 3.2 were used. For "city" categorical feature we have chosen 185 cities with more than 10 citizens from our dataset and preprocessed the feature with one-hot encoding. "Gender" feature in VK has several possible values (male, female, unknown), we present it as two one-hot encoded features. Overall, with one-hot encoding of categorical features we obtained 204 features for each account. 

Account information is usually sparse - accounts have many blank fields. We considered two approaches for empty fields filling. The first approach is to use some constant value. The second - to use average value among the filled field values. With the first approach on all 204 features we obtained ROC AUC 0.785, with the second approach - 0.762.

\subsection{Classification on Node2Vec graph embeddings}

We trained 25 Node2Vec models with different p and q parameters. The grid for each of the parameters is $\{0.25, 0.5, 1, 2, 4\}$ (the same as in the original paper \cite{node2vec}). The other parameters are 10 random walks of length 80 per source, context size - 10 and 10 epochs of SGD optimization.

We consider AUC-ROC (Area Under ROC Curve) as the main metric for our binary classification task.  We train Logistic Regression Classification algorithm based on the obtained Node2Vec embeddings. The results for technical and sophisticated accounts are presented in Table \ref{table:node2vec_grid_1} and \ref{table:node2vec_grid_2}, respectively. 

\begin{table} 
\centering
\small
\begin{tabular}{ c||c|c| c| c|c } 
&p = 0.25 & p = 0.5 & p = 1 & p = 2 & p = 4\\
\hline
\hline
q = 0.25 & 0.856 &	0.814 &	0.804 &	0.823 &	0.780 \\
q = 0.5 & 0.787 & 0.768 & 0.813 & 0.799 & 0.822 \\
q = 1 &	0.863 &	0.812 &	0.847 &	0.829 &	0.808 \\
q = 2 & 0.821 &	\textbf{0.931} &	0.776 &	0.793 &	0.848 \\
q = 4 & 0.793 &	0.801 &	0.848 &	0.809 &	0.823\\
\end{tabular}
\caption{LogReg Classification ROC AUC results for technical accounts based on Node2Vec embeddings with different p and q parameters, number of dimensions - 50.}
\label{table:node2vec_grid_1}
\end{table}

We also considered SVM and Random Forest classification algorithms trained on the same embeddings as in Table \ref{table:node2vec_grid_1}, those approaches best results are 0.929 and 0.922 respectively. The best Logistic Regression results outperform SVM and Random Forest results.

\begin{table}
\centering
\small
\begin{tabular}{ c||c|c| c| c|c } 
&p = 0.25 & p = 0.5 & p = 1 & p = 2 & p = 4\\
\hline
\hline
q = 0.25 & 0.727 & \textbf{0.823} & 0.751 & 0.753 & 0.793 \\
q = 0.5 & 0.750 & 0.795 & 0.796 & 0.806 & 0.754 \\
q = 1 &	0.771 & 0.804 &	0.765 & 0.788 & 0.772 \\
q = 2 & 0.747 & 0.742 & 0.808 & 0.764 & 0.779 \\
q = 4 & 0.776 & 0.724 & 0.745 &  0.709 & 0.793\\

\end{tabular}
\caption{LogReg Classification ROC AUC results for sophisticated accounts based on Node2Vec embeddings with different p and q parameters, number of dimensions - 100.}
\label{table:node2vec_grid_2}
\end{table}

Table 2 (sophisticated accounts) shows a significant decrease in quality compared to Table 1 (technical accounts). Graph random walk parameters are the same. The best model for sophisticated accounts classification shows 0.823 ROC-AUC score.

\subsection{Joint graph structure and account features learning} 

Attri2Vec approach allows to use account features with graph structure features. We used authors implementation from the original paper \cite{attri2vec}. We considered the same length of random walks, number of random walks and context size parameters as for Node2Vec embeddings - 80, 10 and 10, respectively. The other training parameters are 1M samples and 0.025 initial learning rate value.

We considered several types of the mapping options for constructing embeddings from attributes in Attri2Vec:  ReLU mapping, Fourier mapping and Sigmoid mapping. The best results were shown by Sigmoid mapping. The best obtained results for technical and sophisticated accounts are presented in Table \ref{table:attri2vec_datasets}.

\begin{table}[h] 
\centering
\small
\begin{tabular}{ c||c| c } 
&Technical accounts & Sophisticated accounts \\
\hline
\hline
AUC ROC & \textbf{0.988} & 0.684 \\

\end{tabular}
\caption{LogReg Classification ROC AUC results based on Attri2Vec embeddings.}
\label{table:attri2vec_datasets}
\end{table}

The obtained Attri2Vec results outperform Node2Vec results for technical accounts. Results on the sophisticated accounts are worse. This can be caused by more complex structure and profile fields completion of sophisticated accounts and large number of gaps in account features of sophisticated accounts. 

\subsection{Classification on concatenated Node2Vec embeddings and account features} 

We considered classification based on concatenated Node2Vec best embeddings and account features. We used Logistic Regression Classification algorithm. The best results on technical and sophisticated accounts are presented in Table \ref{table:concatenated}.

\begin{table}[h] 
\centering
\small
\begin{tabular}{ c||c| c } 
&Technical accounts & Sophisticated accounts \\
\hline
\hline
AUC ROC & 0.911 & \textbf{0.867} \\
Zegzhda et.al. \cite{zegzhda}  & --- & 0.73  \\
Skorniakov et.al. \cite{ispras} & --- & 0.820 \\
\end{tabular}
\caption{LogReg Classification ROC AUC results based on concatenated Node2Vec embeddings and account features.}
\label{table:concatenated}
\end{table}

Concatenation of Node2Vec embeddings and account features outperforms account features results for both datasets and gives the best obtained results on sophisticated dataset. Graph based methods show the best results on technical accounts dataset, but fail to classify sophisticated accounts without profile features.

\section{Conclusions}
In this work, we proposed simple and advanced sets for the bots detection problem. We showed the effectiveness of using graph embedding and combining it with user attribute features to solve this problem. The best ROC AUC score of 0.988 for technical accounts and 0.867 for sophisticated accounts shows the complexity difference in proposed tracks. Unfortunately, we can’t directly compare our results with existing approaches because most works do not provide the necessary datasets and code repositories. We took the best results of the bot detection from the paper by Zegzhda et.al. \cite{zegzhda} and Skorniakov et.al. \cite{ispras} as most close to our work. Proposed approach outperforms both approaches by 4.5\% of AUC\footnote{We did not re-implement the method and only give the result published by the author}. Achieved result is sufficient for application purposes, greater results can be reached with the following strategies:
\begin{itemize}
    \item use of text embeddings - a significant part of artificial accounts performs the function of promoting certain goods or disseminating information, which can be used for classification;
    \item significant number of accounts hide their friends, but leave open groups that can be used to model a user as a bipartite graph node;
    \item network modeling as a temporal network is of interest, taking into account such characteristics as the joint appearance of accounts on the network.
\end{itemize}

Resulting anonymized dataset, pretrained embeddings and evaluating scripts can be found at project repository. \footnote{Bot detection dataset as well as the project code are available at \url{http://github.com/karpovilia/botdetection}}

\section{Acknowledgements}
This work was supported by Russian Academic Excellence Project `5-100'. We thank Daria Musatkina, Maksim Smirnov and Matvey Osmolovsky for assistance with data processing and meaningful discussion.

\bibliographystyle{plain}
\bibliography{main}
\end{document}